\documentclass[prd,aps,nofootinbib,superscriptaddress,eqsecnum,floatfix,preprintnumbers,amsmath,amssymb,
nofootinbib,longbibliography,twocolumn]{revtex4-2}

\usepackage{amssymb,amsmath}
\usepackage{epsfig}
\usepackage[dvipsnames]{xcolor}
\usepackage[utf8]{inputenc}
\usepackage{stmaryrd}
\usepackage{mathrsfs}
\usepackage{mathalfa}
\usepackage{accents}
\usepackage[normalem]{ulem}

\usepackage{graphicx}

\newcommand{\pa}{\partial}

\newcommand{\be}{\begin{equation}}
\newcommand{\ee}{\end{equation}}
\newcommand{\bea}{\begin{eqnarray}}
\newcommand{\eea}{\end{eqnarray}}
\newcommand{\ba}{\begin{equation}\begin{aligned}}
\newcommand{\ea}{\end{aligned}\end{equation}}

\newcommand{\beg}{\begin{gather*}}
\newcommand{\eng}{\end{gather*}}
\newcommand{\hh}{,\hspace{0.5cm}}
\newcommand{\hhh}{,\hspace{0.3cm}}

\newcommand{\n}[1]{\label{#1}}

\newcommand{\ins}[1]{{\mbox{\tiny #1}}}
\newcommand{\ind}[1]{{\mbox{\scriptsize #1}}}

\def\XXint#1#2#3{{\setbox0=\hbox{$#1{#2#3}{\int}$ }
\vcenter{\hbox{$#2#3$ }}\kern-.6\wd0}}

\usepackage[makeroom]{cancel}
\usepackage[caption=false]{subfig}
\usepackage[colorlinks=true,
            citecolor=green,
            linkcolor=red,
            filecolor=cyan,
            urlcolor=magenta,
            backref=false]{hyperref}

\begin{document}

\title{Regular black holes in quasitopological gravity:\\ Null shells and mass inflation}

\author{Valeri P. Frolov}%
\email[]{vfrolov@ualberta.ca}
\affiliation{Theoretical Physics Institute, Department of Physics,
University of Alberta,\\
Edmonton, Alberta, T6G 2E1, Canada
}

\author{Andrei Zelnikov}%
\email[]{zelnikov@ualberta.ca}
\affiliation{Theoretical Physics Institute, Department of Physics,
University of Alberta,\\
Edmonton, Alberta, T6G 2E1, Canada
}

\begin{abstract}
We investigate the phenomenon of mass inflation in the interior of regular black holes arising in quasitopological gravity (QTG). These geometries are characterized by a bounded curvature core and the presence of an inner (Cauchy) horizon located near the fundamental scale $\ell$. To examine whether mass inflation persists in this setting, we model the interaction of ingoing and outgoing perturbations by considering the collision of two spherical null shells inside the black hole. Using the Dray--'t\,Hooft--Barrabes--Israel junction condition, we derive conditions under which the metric function and curvature invariants may experience significant amplification near the inner horizon. Our analysis shows that, unlike in classical Reissner--Nordstr\"om or Kerr geometries, significant mass inflation requires shell intersection at radii very close to the horizon, with radial separations from it of the order $r-r_* \lesssim \ell \big(\ell/r_\ind{g}\big)^{2n(D-3)}$, where $r_\ind{g}$ is the gravitational radius of the black hole, $D$ is the number of spacetime dimensions and $n\ge 1$ is a parameter depending on a concrete QTG model. For macroscopic black holes with $r_\ind{g}\gg \ell$ this distance is much smaller than the fundamental scale $\ell$. We discuss possible consequences of this effect.

\medskip
\hfill {\scriptsize Alberta Thy 5-25}
\end{abstract}

\maketitle

\section{Introduction}

The interior structure of black holes remains a central subject in both classical and semiclassical gravitational physics. One of the key phenomena that shapes our understanding of this region is \emph{mass inflation} \cite{PoissonIsrael1990}, first recognized in studies of the inner (Cauchy) horizon of Reissner-Nordström and Kerr black holes
\cite{PoissonIsrael1990, Ori1991, MarkovicPoisson1995,BradySmith1995, Burko:1997zy,Burko:1999zv,OrenPiran2003, Dafermos2005, HamiltonPollack2005,HamiltonAvelinoReview2010,Carballo-Rubio:2018jzw}.
(see also books \cite{FrolovNovikov1998,Poisson2004} and references therein).

\begin{figure}[hbt]
    \centering
   \includegraphics[width=0.32
      \textwidth]{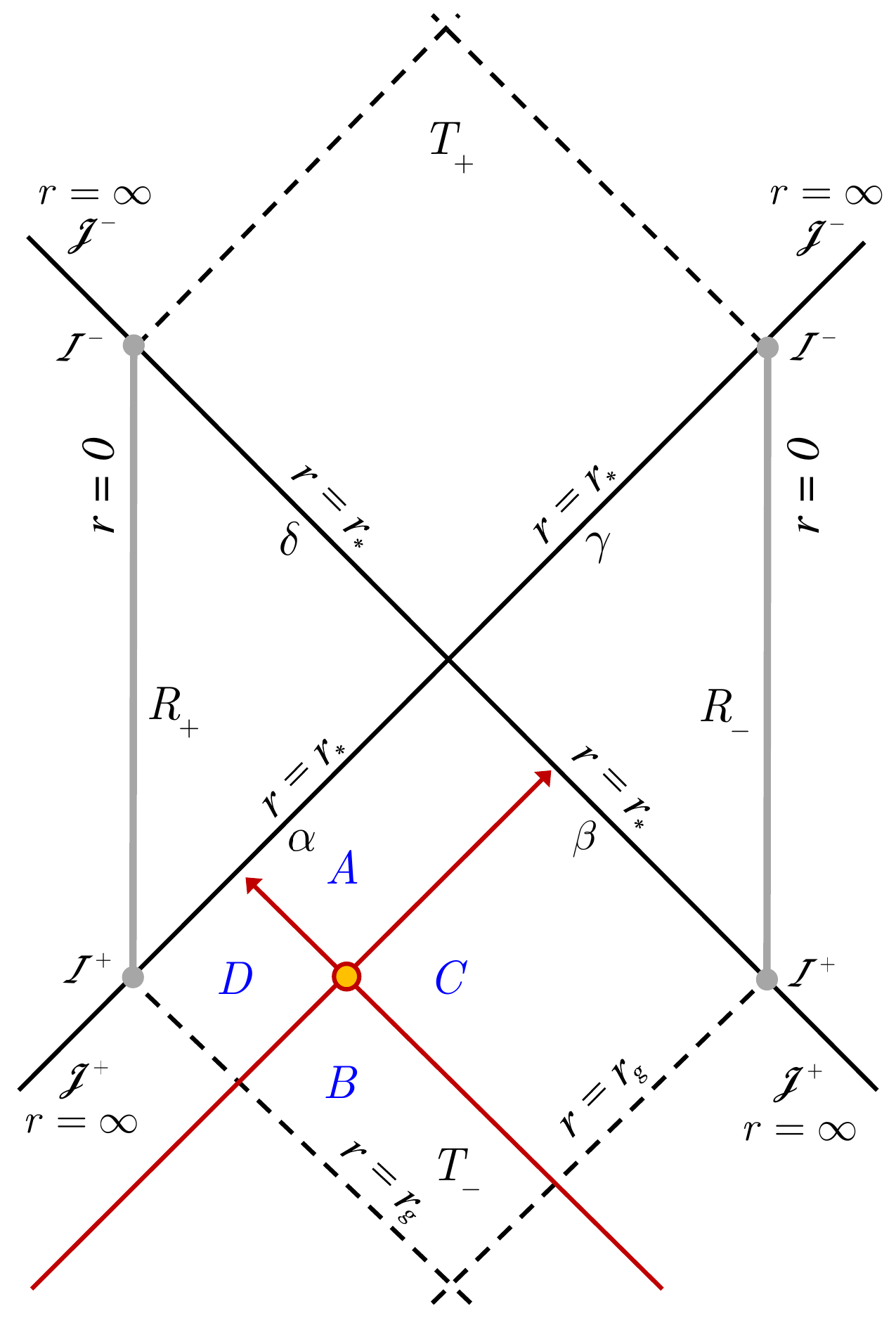}
    \caption{Colliding spherical thin null shells near the inner (Cauchy) horizon inside the black hole. The domain $T_-$ splits into four regions $A,B,C,$ and $D$, separated by null shells.
    }
    \label{Penrose_3}
\end{figure}

To explain the mechanism of the mass inflation
let us consider a
charged (Reissner--Nordstr\"om) black hole and assume that inside it there exist  two families of lightlike matter streams:
\begin{itemize}
    \item ingoing null radiation moving toward the inner (Cauchy) horizon and
    \item outgoing null radiation that was emitted earlier and is now propagating outward inside the black hole.
\end{itemize}
Each stream can be idealized as a sequence of infinitesimally thin null shells. After an ingoing and outgoing null shells intersect their masses change.
This process is illustrated by
Fig.~\ref{Penrose_3}.

There exist four spacetime domains bounded by null shells, which are denoted as $A$, $B$, $C$, and $D$.
In each of these domains
the spherically symmetric  metric has the form
\be
ds^2=-f(r)dt^2 +\dfrac{dr^2}{f(r)}+r^2 d\omega^2\, ,
\ee
where $d\omega^2$ is a metric of a unit $2D$-sphere.
The metric functions in each of the domains depend on the mass parameters, which we denote by $m_A$, $m_B$, $m_C$, and $m_D$ respectively.
Even if each of the colliding shells carries small energy, the change becomes significant when the collision happens near the inner (Cauchy) horizon.

There exists a remarkable simple matching condition at the intersection sphere of the null shells which has the form
\be \n{ABCD}
f_A=\dfrac{f_C f_D}{f_B}\ .
\ee
Here all the metric functions are taken at the radius $r_i$ of the intersection sphere. This condition was derived in the papers by 't\,Hooft and Dray \cite{tHooft:1985} and Barrabes and Israel \cite{BarrabesIsrael1991}. In what follows we refer to \eqref{ABCD} as the DHBI condition.
This condition implies that the effective internal mass parameters change. It should be emphasized that  the DHBI condition is proved under quite general conditions and does not require the Einstein equations.  It remains valid for quite general class of modified gravity theories. In this sense the DHBI condition is universal.

Near the inner horizon where $f_A$ vanishes, ingoing and outgoing radiation experience extreme mutual blueshift. To an infalling observer, outgoing radiation appears with increasing energy; to outgoing radiation, the infalling flux appears amplified. This feedback transforms small initial perturbations into large gravitational effects.
This phenomenon is known as \emph{mass inflation}.

As a result of the mass inflation the inner (Cauchy) horizon is generically unstable once even a small amount of radiation is present. Mass inflation arises because ingoing and outgoing null radiation mutually blueshift near the inner horizon; crossings of null shells convert this effect into an exponential growth of the internal mass parameter.

Certainly, the crossing thin null shell model is oversimplified. However, further study showed that the main feature of the mass inflation effect still exists when ingoing and outgoing distributed fluxes interact in the near-horizon region. The strong blueshift of perturbations can lead to a rapid growth of the effective internal mass parameter.
Mass inflation has been explored through several complementary approaches, including perturbative analyses, numerical relativity, and effective models that incorporate radiative backreaction. Foundational work by Poisson and Israel \cite{Poisson:1989zz,Poisson:1990eh} demonstrated that the focusing of radiation near the Cauchy horizon naturally leads to exponential growth of curvature. Subsequent research extended this picture to regular black holes, modified theories of gravity, and quantum-corrected geometries. This behavior indicates that the naive classical continuation of the interior geometry breaks down, raising questions about predictability and about the physical status of the inner horizon.
A simplified model of intersecting thin null shells captures important features of the mass inflation and can  serve as an important test of the mass inflation effect.
A central issue in these developments is whether mass inflation is an unavoidable feature of black hole interiors, or whether certain models admit nonsingular behavior and a stable inner horizon.

In the present paper, we reconsider the problem in the context of regular black hole models. Namely, we consider regular black holes solutions of a so-called
\emph{quasitopological gravity theories} (QTG)
\cite{Oliva_2010,Myers:2010ru, Bueno:2016cu, Hennigar:2017ego, Bueno:2019ltp, Bueno:2022res, Moreno:2023rfl, Hao:2025utc}.
In this framework, the standard Einstein-Hilbert action is extended by an infinite tower of higher-curvature contributions. These additional terms consist of polynomial scalar invariants formed from the curvature, modifying the theory at sufficiently high energies. For a $D$-dimensional spacetime, the action can be expressed schematically as
\begin{equation}\n{WWW0}
W_{QTG}=\frac{1}{2\kappa}\int d^D x\,\sqrt{-g}\; L(g,R)\, ,
\end{equation}
with the Lagrangian density
\begin{equation}\n{WWW1}
L(g,R)=R+\sum_{n=2}^{\infty}\alpha_n\,\ell^{2(n-1)}\,\mathcal{Z}_n\, .
\end{equation}
Here $\mathcal{Z}_n$ represents the complete collection of independent scalar invariants of order $n$ that can be constructed from contractions of the Riemann tensor. The coefficients $\alpha_n$ are dimensionless constants determining the strength of each higher-curvature contribution, while $\ell$ is a characteristic length scale that governs the onset of deviations from classical general relativity by setting the curvature threshold at which these corrections become non-negligible.

In the generic situation, the field equations for the action \eqref{WWW0}-\eqref{WWW1}
contain derivatives of the metric up to fourth order, which arise from variations of
the terms $\mathcal{Z}_n$. Nevertheless, it is known
that for spacetime dimension $D \ge 5$ and for every $n \ge 2$, one can
select the polynomial curvature invariant $\mathcal{Z}_n$ so that the
corresponding Euler-Lagrange equations remain at most
second order when evaluated on a spherically symmetric metric. Thus,
even though the action includes higher-curvature contributions, the
resulting field equations preserve their second-order form within the
spherically symmetric sector, mirroring the structure familiar from
classical general relativity.
After imposing spherical symmetry, the reduced action attains a closed analytic form, which makes the resulting field equations accessible without relying solely on perturbative expansions.

When the series of higher-curvature contributions does not truncate, and the associated coefficients $\alpha_n$ satisfy the necessary consistency conditions, the equations obtained from the reduced action admit solutions describing \emph{regular black holes}. In this setting, regularity refers to the absence of curvature singularities: curvature scalars remain finite at the center, and the geometry avoids the conventional divergence encountered in standard black hole interiors.

Recent work has shown that such solutions occur broadly across many realizations of quasitopological gravity (see, for instance,
\cite{Bueno:2024dgm, Frolov:2024hhe, Bueno:2025qjk, Bueno:2025zaj, Ling:2025ncw,
Sueto:2025ufn, Bueno:2025gjg, PinedoSoto:2025hel} and related discussions). These studies collectively indicate that regular black holes arise systematically rather than as isolated or fine-tuned special cases.
These QTG models contain a characteristic length scale $\ell$, which plays the role of a fundamental cutoff. For spherically symmetric regular black holes in this framework, an inner horizon typically forms at a radius close to $r=\ell$. A generic feature of such solutions is the appearance of a de~Sitter-like core, where curvature scalars remain finite and polynomial invariants are bounded.

Our goal is to analyze whether mass inflation can develop near this inner horizon of a regular black hole solution of QTG\footnote{For a related discussion of this problem see also \cite{Carballo-Rubio:2018jzw,DiFilippo:2024mwm}.
}. To this end, we study a simplified model involving two spherical null shells in the black hole interior intersecting just inside the inner horizon. We assume that the gravitational radius $r_\ind{g}$ of the black hole is much larger that the fundamental length $\ell$, so that the dimensionless parameter $\beta=\ell/r_\ind{g}$ is small.
Then
by applying the DHBI junction conditions, we find that a significant inflationary effect on the geometry occurs only when the intersection radius $r$ is close to the inner horizon $r=r_*$
\be
\Delta r=r-r_* \sim \frac{ \Delta m_C \, \Delta m_D }{m^2}\,\Big(\dfrac{\ell}{r_\ind{g}}\Big)^{2n(D-3)}\, \ell .
\ee
Here $D$ is the number of spacetime dimensions, the parameter $n\ge 1$ depends on the concrete version of QTG, $\Delta m_C$ and $\Delta m_D$ are the mass parameters of the colliding shells, $m$ is the background black hole mass parameter, and $r=r_*$ is the radius of the inner horizon.
This condition differs substantially from the corresponding behavior in the classical charged or rotating solutions of Einstein gravity.
This result implies that for small value of the parameter $\beta$ the radial distance $\Delta r$ from the inner horizon is much smaller than the fundamental scale $\ell$.
This observation has implications for the internal consistency of quasitopological gravity theories and for the broader question of inner horizon stability.

The paper is organized as follows. In Sec.\,II we discuss regular black hole solutions of the QTG. In Sec.\,III we use the model of intersecting spherical null shells and use the DHBI junction relation for derivation of the mass inflation condition for two special models of QTG regular black hole. In Sec.\,IV we derive a similar condition for a wide class of QTG regular black holes. Section~V contains a discussion of the obtained results and their possible generalizations and application. We use the signature conventions of Ref.\cite{MTW}.

\section{Regular black holes in QTG}

\subsection{Spherically reduced action}

Let us consider a $D$-dimensional static spherically symmetric spacetime with the metric of the form
\be \n{MET}
ds^2=-N^2(r) \, dt^2 f(r)+\dfrac{dr^2}{f(r)}+r^2 d\Omega_{D-2}^2\, .
\ee
For this metric the reduced action\footnote{The integral over $t$ is understood as an interval $\Delta t$ of time between some initial $t_1$ and final $t_2$ moments of time, $\Delta t=t_2-t_1$. This constant factor does not affect the field equations.} of the QTG model is
\be \n{WWR}
W_{red}=B\int dt dr N(r) \big(r^{D-1}h(p)\big)'\, .
\ee
Here and later a prime denotes a derivative over $r$, and
\be
B=\dfrac{(D-2)\Omega_{D-2}}{2\kappa}\hh  \kappa=8\pi G^{(D)}\, .
\ee
where $G^{(D)}$ is a $D$-dimensional Newton's coupling constant.
We also denote
\be \n{pp}
p=\dfrac{1-f}{r^2}\, .
\ee
This is one of the four basic curvature invariants of a spherically symmetric static metric (see e.g. \cite{Frolov:2024hhe}). We call it primary basic curvature invariant, or simply {\it primary invariant}.
In the QTG the reduced action \eqref{WWR} is obtained for $D\ge 5$. We shall not impose this constraint and use \eqref{WWR} as an effective reduced action for some $4D$ theory as well.

The function $h=h(p)$, which enters the reduced action \eqref{WWR}, is specified by a chosen QTG model. We do not fix it at the moment but simply assume that it is chosen so that the corresponding solutions of the QTG describe regular black holes. Two special  examples of such solutions will  be considered in the next section.
The field equations obtained by varying of the reduced action \eqref{WWR} with respect to its arguments $f$ and $N$ give
\be \n{NNpp}
N=\mbox{const}\hh h(p)=\dfrac{m}{r^{D-1}}\, .
\ee
We put $N=1$. The constant $m$ is related with the mass $M$ of the black hole
\be\n{mmm}
m=\dfrac{2\kappa M}{(D-2)\Omega_{D-2}}\, .
\ee

\subsection{Regular black holes}

For $N=1$ the metric \eqref{MET} is
\be\n{metric}
\begin{split}
ds^2&=-fdt^2+\dfrac{dr^2}{f}+r^2 d\Omega_{D-2}^2\, ,\\
f(r)&=1-r^2 p(r)\, ,
\end{split}
\ee
where  $p=p(r)$ is the primary curvature invariant; it is recovered by inverting the functional relation between
$h(p)$ and $p$ given in
\eqref{NNpp}.
In what follows we assume that $p(r)$ is a bounded function on the interval $r\in [0,\infty)$ and it has the following asymptotic form when $r\to\infty$
\be \n{ass}
p(r)\approx \dfrac{m}{r^{D-1}}\, .
\ee
For $h(p)=p$ one has the Schwarzschild-Tangherlini metric and the mass parameter is related to the gravitational radius of this black hole $r_\ind{g}$ as follows
\be \n{rg}
m=r_\ind{g}^{D-3}\, .
\ee
The metric \eqref{MET} has three independent basic curvature invariants which we denote $p$, $q$ and $v$. The invariants $q$ and $v$ can be expressed in terms of $p(r)$ and its derivatives
\be
q=\dfrac{1}{2}(rp'+2p)\hh
v=r^2p''+4rp'+2p\, .
\ee
All polynomial scalar curvature invariants for the metric \eqref{MET} can be written as polynomial functions of its basic invariants $p$, $q$ and $v$. If the primary basic invariant $p$ is bounded, then the other basic invariants, as well as their polynomials are bounded as well.

At the points where $f(r)=0$ the solution has a horizon. Since both at the infinity and at the center the function $f(r)$ takes the same value 1, the number of the horizons is even. We assume that this number is equal to 2. We call such a solution with bounded $p(r)$ satisfying condition \eqref{ass} a {\it regular black hole}.

\begin{figure}[hbt]
    \centering
      \includegraphics[width=0.35
      \textwidth]{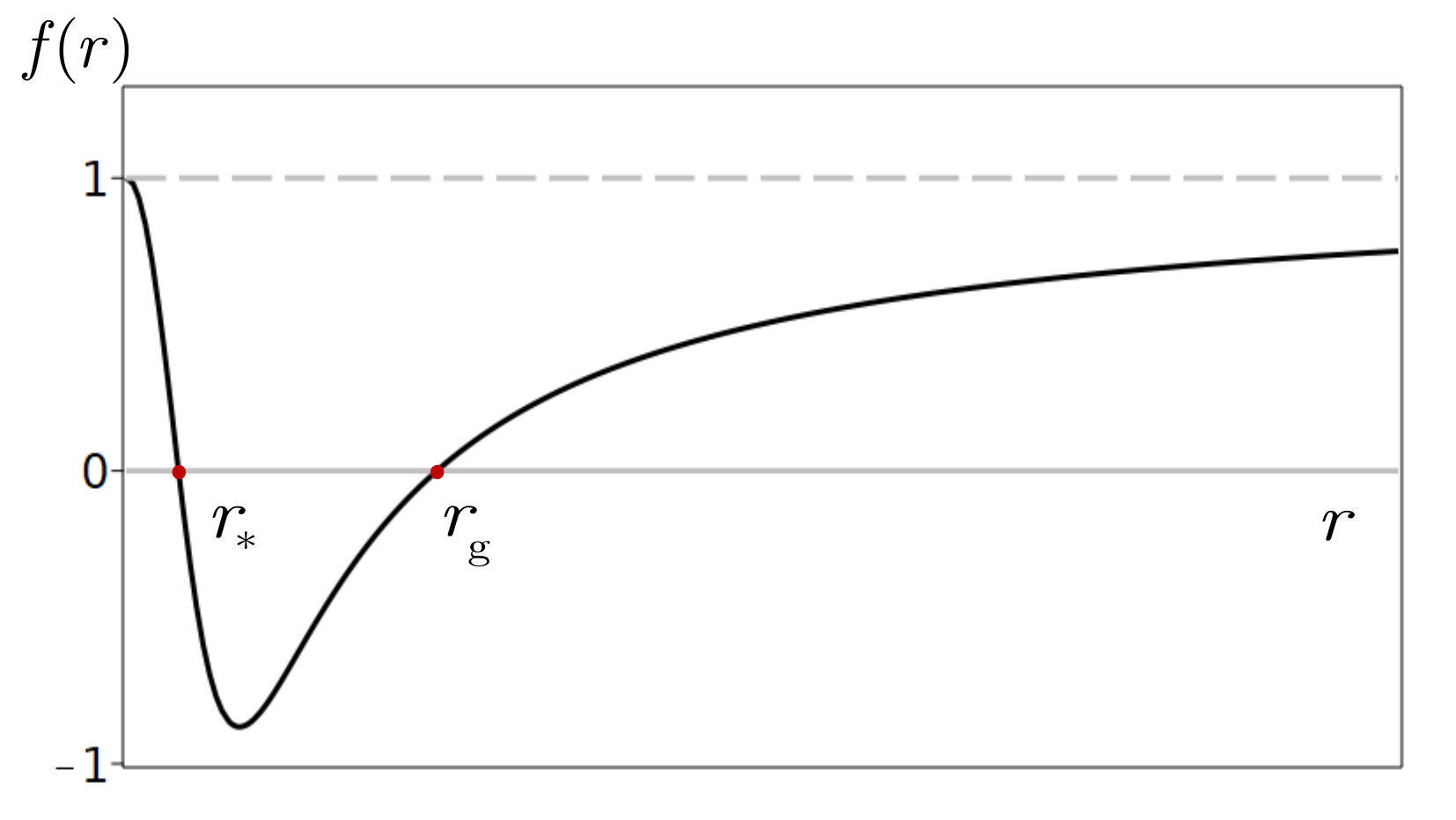}
    \caption{The typical shape of functions $f(r)$ for regular black holes. The external horizon is located at $r_\ind{g}$ and the internal one is at $r_*.$}
    \label{f_Plot_1}
\end{figure}

A characteristic property of regular black holes is the existence of the mass gap $m_0$, such that for mass parameter $m<m_0$ the metric does not have horizons.
For  $m>m_0$ there exist the external (event)  and internal (Cauchy) horizons (see Fig.~\ref{f_Plot_1}).


\section{"Mass inflation" in QTG models }

Let us consider now  a static spherically symmetric metric spacetime describing a regular black hole. We assume that there exist two thin null shells, one incoming and the other outgoing, which intersect in the domain inside the black hole as shown in  Fig.~\ref{Penrose_3}.

To make a consideration more concrete in this section we first discuss two special QTG solutions: one which describes Hayward-type regular black hole and the other with Born-Infeld regular black hole solution.

\subsection{"Mass inflation" in Hayward-type black hole}

As a first example, we consider a $D$-dimensional Hayward-type solution of the QTG equations. For this model
\be
h(p)=\dfrac{p}{1-\ell^2 p}\, .
\ee
Inverting this equation gives
\be
p=\dfrac{h}{1+\ell^2 h}\, .
\ee
Here $\ell$ is length scale parameter. Substituting expression for $h$ from \eqref{NNpp}
\be
h=\dfrac{m}{r^{D-1}}\,
\ee
one finds
\be\n{HH}
\begin{split}
p&=\dfrac{m}{r^{D-1}+\ell^2 m}\, ,\\
f&=1-\dfrac{m r^2}{r^{D-1}+\ell^2 m}\,.
\end{split}
\ee
The corresponding metric is nothing but a $D$-dimensional version of the Hayward metric.

In the limit  $\ell=0$ one has $h(p)=p$ and the corresponding metric
\be
f=1-\dfrac{m}{r^{D-3}}
\ee
reproduces the Schwarzschild-Tangherlini solution of the Einstein equations.
Denote
\be \n{mgr}
m=r_\ind{g}^{D-3}\, .
\ee
Then $r=r_\ind{g}$
is the radius of the event horizon of such black hole.
Using this parameter it is convenient to present the primary basic curvature $p(r)$ and the metric function $f(r)$ for the Hayward-type metric \eqref{HH} in the form
\be \n{ppff}
p=\dfrac{1}{\ell^2}\dfrac{1}{1+\beta \rho^{D-1}}\hh
f=1-\dfrac{\rho^2}{1+\beta \rho^{D-1}}\, ,
\ee
where
\be \n{rh}
\rho=\frac{r}{\ell} \hh
\beta=\Big(\dfrac{\ell}{r_\ind{g}}\Big)^{D-3}\, .
\ee

The function $f=f(r)$ has zeros only if the mass parameter $m\ge m_0$.
The mass gap $m_0$ is defined as a solution of the following two equations:
\be
f=0  \hh f'=0\, .
\ee
One finds the value of the inner horizon radius $r_*=\ell \rho_0$ of the regular black hole with the minimal mass  and the maximal inverse mass parameter $\beta_0=\ell^{D-3}/m_0$ for the black hole to exist:
\be
\rho_0=\sqrt{\dfrac{D+1}{D-1}}\hh
\beta_0=\dfrac{2}{D-1}\Big( \dfrac{D-1}{D+1}\Big)^{(D-3)/2}\, .
\ee

In what follows we assume that the gravitational radius $r_\ind{g}$  is much larger than the fundamental length $\ell$ and the parameter $m_0$.
This means that  $\beta$ is a small parameter.

Solving equation $f(r)=0$ one finds the following expression the dimensionless radius of the inner horizon
\be
\rho_{*}=1+\dfrac{1}{2}\beta+\frac{2D-3}{8}\beta^2+O(\beta^3)\, .
\ee
In what follows it is sufficient to keep only first three terms of this expansion.

Consider a point close to the inner horizon $\rho=\rho_*$\,:
\be
\rho=\rho_*+x\hh x>0, \ \ x\ll 1\, .
\ee
Then  one has
\ba\n{fB}
f=&-2\Big[1-\frac{D}{2}\beta-\dfrac{2D^2-8 D+3}{8}\beta^2+O(\beta^3)\Big] x\\
&+O(x^2)\, .
\ea

Let us consider now situation when there exist two spherical null shells propagating in the black hole interior (see Fig.~\ref{Penrose_3}). We assume that they intersect at some radius $r_i$ close to the inner horizon of the metric in region $B$. We assume that the mass parameter in this domain is $m$ and use the relation \eqref{fB} for the approximate value of $f_B$ at the point of the shells' intersection.

We write the mass parameters $m_C$ and $m_D$ in domains $C$ and $D$ as follows
\be
m_C=m/(1-\mu_C)\hh m_D=m/(1-\mu_D)\, .
\ee
\be
\mu_C=\frac{m_\ins{in}}{m_B+m_\ins{in}},\hskip 0.8cm \mu_D=\frac{m_\ins{out}}{m_B+m_\ins{out}} \, .
\ee
Then metric functions in $C$ and $D$ are given by \eqref{ppff} where
\be
\beta\to \beta_C=\beta(1-\mu_C)\hh
\beta\to \beta_D=\beta(1-\mu_D)\, .
\ee
It is easy to check that keeping the leading terms for small $x$ and $\beta$ one finds
\be
\begin{split}
f_C&=-\beta \mu_C(1+x)+\cdots\, ,\\
f_D&=-\beta \mu_D(1+x)+\cdots \, .
\end{split}
\ee
Thus one has
\be\label{f_A}
f_A=\dfrac{f_C f_D}{f_B}\approx
-\beta^2\dfrac{\mu_C \mu_D }{2 x}\, .
\ee
This result implies that if the radius of the intersection sphere approaches the horizon arbitrarily closely, the metric function $f_A$ becomes arbitrarily large. In Einstein gravity such behavior corresponds to an unbounded growth of the effective mass parameter, the phenomenon known as mass inflation. It is therefore convenient to adopt the condition $|f_A|\sim 1$ as a practical criterion for the onset of mass inflation. In the near-horizon region the metric function $|f|$ prior to the shell intersection is very small. A change in its magnitude to a value of order unity after the collision corresponds to an “explosive’’ modification of the metric. This criterion will be used throughout the remainder of the paper.

Relation (\ref{rr_*}) shows that, for a regular black hole, achieving the condition $|f_A|\sim 1$ requires the intersection radius of the shells to be extremely close to the horizon of the $B$ region:
\be \n{rr_*}
r-r_*=x\sim \mu_C \mu_D \Big(\dfrac{\ell}{r_\ind{g}}\Big)^{2(D-3)} \ell.
\ee
If $\ell$ is the Planck length, then closeness should be deep inside the trans-Planckian domain. This also imposes a restriction on the thickness of null shells, which is obviously unphysical.

\subsection{"Mass inflation" in  regular Born-Infeld black hole}

As a second example, we consider now
a regular Born-Infeld black hole. Its metric is a solution of the QTG equations for the following choice of the function $h(p)$
\be
h=\dfrac{p}{\sqrt{1-\ell^4 p^2}}\, .
\ee
Inverting this relation one gets
\be
p=\dfrac{h}{\sqrt{1+\ell^4 h^2}}\, .
\ee
After substituting \eqref{NNpp} one finds
\be
p=\dfrac{m}{\sqrt{r^{2(D-1)}+\ell^4 m^2}}\, .
\ee
It is convenient to introduce dimensionless quantities similar to \eqref{rh}:
\be \n{rh1}
\rho=\frac{r}{\ell}\hh
\beta=\Big(\dfrac{\ell}{r_\ind{g}}\Big)^{D-3}\, .
\ee
Then one obtains
\be \n{fA}
f_B=1-\frac{\rho^2}{\sqrt{1+\beta^2 \rho^{2D-2}}}\, ,
\ee
where $\beta$ corresponds to mass parameter $m$ of the black hole in the domain $B$
\be
m=r_\ind{g}^{D-3} .
\ee

We assume that $\beta\ll 1$.
$f_B=0$ at the inner horizon $\rho=\rho_*$ where
\be
\rho_{*}=1+\dfrac{1}{4}\beta^2+\frac{4D-7}{32}\beta^4+O(\beta^6)\, .
\ee

Further we will keep in all expression terms up to the second order in $\beta$ only.

Denote
\be
\rho=\rho_{*} +x\, .
\ee
Then for $x\ll 1$ one has
\be
f_B=-2\Big[1-\frac{2D-1}{4}\beta^2+O(\beta^4)\Big]x+O(x^2)\, .
\ee
Hence $x=0$ is a position of the horizon for $f_B$.

Let us assume that in domains $C$ and $D$ one has
\be
m_C=m_B/(1-\mu_C)\hh m_D=m_B/(1-\mu_D)\, ,
\ee
\be
\mu_C=\frac{m_\ins{in}}{m_B+m_\ins{in}} \hh  \mu_D=\frac{m_\ins{out}}{m_B+m_\ins{out}} \, .
\ee
Then the metric functions in $C$ and $D$ are given by \eqref{fA} where
\be
\beta\to \beta_C=\beta(1-\mu_C)\, ,
\mbox{\ and \ }
\beta\to \beta_D=\beta(1-\mu_D)\, .
\ee
Calculating $f_A$ according to (\ref{ABCD}) we obtain the asymptotics at small $x$:
\be
f_A\approx-\beta^4\frac{\mu_C \mu_D(2-\mu_C)(2-\mu_D) }{8 x}+O(1)
\, .
\ee
The corrections  $O(1)$  are finite and proportional to $\beta^2$. In order $f_A$ to become of order 1, the radius $r$ of the shells' intersection sphere  should be extremely close to the horizon of the $B$ domain. For small $\mu_C$ and $\mu_D$
\be \n{rr_*1}
r-r_*=x \sim \mu_C\mu_D\Big(\dfrac{\ell}{r_\ind{g}}\Big)^{4(D-3)} \ell \, .
\ee
If $\ell$ is the Planck length, then closeness should be deep inside trans-Planckian domain. This also imposes a restriction on the thickness of null shells, which is obviously unphysical.

\section{Mass inflation in regular QTG black holes: General case}

In this section we consider again intersection of two spherical null shells. We assume that intersection sphere has radius $r=r_i$ close to the inner horizon.
As earlier we denote by
$A$, $B$, $C$ and $D$ four spacetime domains as it is shown in Fig.~\ref{Penrose_3}. We denote the metric functions in these domains by $f(r)$ with the index corresponding to a chosen domain.

We write the metric function $f$ in the form
\be\n{ffmm}
f(r,m)=1-r^2 p(r,m)\, .
\ee
Here $m$ is mass parameter \eqref{mmm} for a chosen domain
\be
m=r_\ind{g}^{D-3}\, .
\ee
For a regular black hole of mass $m$ the radius $r_*$  of the inner horizon is determined by the equation
\be
r_*^2 p(r_*^2)=1\, .
\ee
In the examples of the QTG regular black holes analyzed in the previous section it was shown that in the limit $m\to \infty$ the radius of the horizon $r_*$ has a well-defined value (which was equal to $\ell$). We assume that this limit exists also in the class of regular black hole models which we discuss now. We denote it by $r_0$. The value of $r_0$ depends on the scale parameter $\ell$ of the chosen QTG model, and for models discussed in the previous sections it coincides with $\ell$.

We introduce the following dimensionless quantities
\be
\rho=\frac{r}{r_0}\hhh \hat{p}=r_0^2 p \hhh
\beta =\Big(\dfrac{r_0}{r_\ind{g}}\Big)^{D-3}\, .
\ee
The dimensionless radius of the inner horizon $\rho=\rho_*$ is defined by the equation
\be \n{rrr}
\rho_*^2\hat{p}(\rho_*,\beta)=1\, .
\ee
According to our assumption
\be
\rho_*|_{\beta=0}=1\hh
\hat{p}(1,0)=1\, .
\ee
Equation \eqref{rrr} determines the dimensionless radius of the inner horizon as a function of $\beta$, $\rho_*=\rho_*(\beta)$.
We assume that
for small $\beta$ this function has the form
\be \n{rrbb}
\rho_*(\beta)=1+b\beta^n +\cdots ,
\ee
where $b$ is a positive dimensionless parameter and $n\ge 1$.

We define
\be
\begin{split}
\kappa&=\kappa(\rho,\beta)=\dfrac{1}{2}\dfrac{\pa f(\rho,\beta)}{\pa \rho}\, ,\\
\kappa_*&=\kappa(\rho_*,\beta)
\, .
\end{split}
\ee
The quantity $\kappa_*$ is (negative) dimensionless surface gravity of the inner horizon.

Differentiation of the relation \eqref{rrr} over $\beta$ gives
\be \n{bbpp}
\rho_*^2 \dfrac{\pa \hat{p}}{\pa \beta}\Big|_{\rho=\rho_*}=2\kappa_* \dfrac{d\rho_*}{d\beta}\, .
\ee
This relation can be used to get expansion \eqref{rrbb} for the solution $\hat{p}(\rho,\beta)$
of the corresponding QTG model.

We assume that $m$ is the mass parameter for $B$ domain, while for $C$ and $D$ domains these mass parameters are
\be\n{DDMM}
m_C=m/(1-\mu_C)\hh
m_D=m/(1-\mu_D) \, .
\ee
The dimensionless parameters $\mu_C$ and $\mu_D$ are connected with the masses $\Delta m_C=m_C-m$ and  $\Delta m_D=m_D-m$ of the intersecting null shell as follows:
\be
\Delta m_C=\dfrac{\mu_C}{1-\mu_C}m\hh
\Delta m_D=\dfrac{\mu_D}{1-\mu_D}m\, .
\ee
We assume that both parameters $\mu_C$ and $\mu_D$ are small, so that the geometry in $C$ and $D$ domains is only slightly different from the geometry in $B$ domain.
Thus we can write
\be \n{ffCC}
\begin{split}
f_C&=f(\rho,\beta-\mu_C\beta)\approx
f(\rho,\beta)-\beta\mu_C\dfrac{\pa f}{\pa \beta}\\
&=f_B-\beta \mu_C\dfrac{\pa f}{\pa \beta}\, ,\\
f_D&=f(\rho,\beta-\mu_D\beta)\approx
f(\rho,\beta)-\beta\mu_D\dfrac{\pa f}{\pa \beta}\\
&=f_B-\beta \mu_D\dfrac{\pa f}{\pa \beta}\,
\, .
\end{split}
\ee

Denote by $x$ a radial distance of the intersection sphere from the horizon of $A$ domain. Then
\be
f_B=f(\rho_*+x,\beta)\approx
2\kappa_* x\, .
\ee
We used here the definition of the inner horizon
\be
f(\rho_*,\beta)=0\, .
\ee
The function $f_A$ at the intersection sphere can be obtained by using DHBI relation
\be
f_A=\dfrac{f_C f_D}{f_B}\, .
\ee
The functions $f_C$ and $f_D$ remain finite at $x\to 0$.
Relations \eqref{ffCC} imply
\be
\begin{split}
f_C&\approx \beta \mu_C \rho_*^2\dfrac{\pa \hat{p}(\rho,\beta)}{\pa \beta}\Big|_{\rho=\rho_*}\approx \beta \mu_C \kappa_*bn \beta^{n-1} \, ,\\
f_D&\approx \beta \mu_D \rho_*^2\dfrac{\pa \hat{p}(\rho,\beta)}{\pa \beta}\Big|_{\rho=\rho_*} \approx \beta \mu_D \kappa_*bn \beta^{n-1}\, .
\end{split}
\ee
Here we used the relation \eqref{bbpp}.

By combining these results we see that the leading (formally divergent) term of the metric function $f_A$ at small $x$ is of the form
\be \n{ffAA}
f_A\approx \dfrac{1}{2}\kappa_0 b^2 n^2 \mu_C \mu_D \dfrac{\beta^{2n}}{x}\, .
\ee
Here we substituted the value of the surface gravity $\kappa_0$ for $\kappa_*$, assuming that for small $\beta$ the difference between these two objects is small.
It is easy to check that \eqref{ffAA} correctly reproduces the results for the Hayward-type black hole (where $r_0=\ell$, $n=1$ and $b=1/2$) and for the Born-Infeld black hole (where $r_0=\ell$, $n=2$ and $b=1/4$.

Relation \eqref{ffAA} shows that
the metric function $f_A$
(and hence the primary basic curvature) formally diverges in the limit $x\to 0$ (mass inflation).
The value of $|f_A|$ becomes of order of one when\footnote{
Let us note that
the proper distance  of the shell intersection sphere to the horizon differs from the radial distance and is of the order of $\, \ell bn\sqrt{\mu_c\mu_D}(r_0/r_\ind{g})^{n(D-3)}$.
}
\be\n{xxxx}
x\sim \dfrac{1}{2}\kappa_0 b^2 n^2 \mu_C \mu_D \Big(\dfrac{r_0}{r_\ind{g}}\Big)^{2n(D-3)}\, .
\ee
Under natural assumptions one can choose
\be
r_0\approx \ell\hh \kappa_0\approx 1\, .
\ee
If we also assume that the parameter $b$ is of the order of 1, then  one arrives to the following relation
\be \n{INEQ}
r-r_* \lesssim \ell \mu_C \mu_D\Big(\dfrac{\ell}{r_\ind{g}}\Big)^{2n(D-3)}\, .
\ee

Let us compare the result \eqref{INEQ} with a similar expression for the mass inflation in the Reissner-Nordstr\"{o}m black hole. For the four-dimensional spacetime the corresponding relation given in the Appendix, \eqref{RNrr}, can be written as follows:
\be\n{RRNN}
r-r_*\approx r_\ind{g} \mu_C \mu_D \dfrac{1}{\sqrt{1-Q^2/M^2}}\, ,
\ee
where
\be
\mu_C=\dfrac{\Delta M_C}{M}\hh
\mu_D=\dfrac{\Delta M_G}{M}\
\ee
for small masses of the intersecting shells have the same meaning as quantities in \eqref{DDMM}.
Note that the last term in \eqref{RRNN} is larger than 1.
This demonstrates that there is a big difference between length scales at which the mass inflation takes place for QTG and the general relativity (in $4D$)
\be \n{llsim}
\dfrac{(r-r_*)_{QTG}}{(r-r_*)_{RN}}\lesssim \Big( \dfrac{\ell}{r_\ind{g}}
\Big)^{2n+1}\, .
\ee

Let us emphasize that the parameter $\ell$ entering \eqref{WWW1} is, {\it a priori}, an arbitrary length scale. However, a natural option is to assume that the higher in curvature terms of the action become important where the curvature becomes close to the Planckian one or slightly smaller than it. For this choice of $\ell$ relation \eqref{INEQ} for macroscopic black holes implies that the effect connected with mass inflation might be important only when the radial distance of the intersection point from the horizon is much smaller than the Planckian length. Let us also mention, that for the validity of the obtained restriction the thickness of the null shells should be much smaller than the Planckian length. Both of these
assumptions seam to be highly unphysical. The QTG model uses the standard classical metric description of the gravity, which certainly is not valid in the trans-Planckian domain, in particular because of possible large fluctuations of the metric.

To summarize, we can conclude that in the QTG models of macroscopic regular black holes  with $\ell\gtrsim \ell_{Pl}$ the mass inflation problem does not arise at least in sub-Planckian domain where the metric $g_{\mu\nu}$, which enters the
action \eqref{WWW0}, is well defined as the classical field.

Let us emphasize, that the above presented derivation does not contain explicit dependence on spacetime dimension $D$, and in this sense it is quite general.

\section{Discussion}

The results obtained in this work clarify the status of mass inflation in regular black holes arising in quasitopological gravity. Using a model of colliding thin null shells, we have shown that, at least formally, the metric function $f(r)$ can grow without bound when the shell-crossing point approaches the horizon. This observation suggests that mass inflation is not directly tied to the regularity of black hole solutions in QTG.

At the same time, the behavior of the models considered here differs substantially from that in classical general relativity. In Reissner-Nordström and Kerr black holes, perturbations near the Cauchy horizon generically lead to an exponential growth of the effective mass parameter. By contrast, the QTG models studied here exhibit a considerably milder growth, indicating that the mechanism of mass inflation is qualitatively modified in this class of regular black hole solutions.

We have demonstrated that if the fundamental length is equal to, or only slightly larger than the Planck length $\ell \gtrsim \ell_{\rm Pl}$, then mass inflation in macroscopic regular black holes within QTG becomes significant only when the intersection of the null shells occurs at a distance from the inner horizon that is much smaller than the Planck scale. In this regime the classical description of mass inflation ceases to be reliable. Such a condition cannot be realized for astrophysical, or even moderately large, black holes. This behavior contrasts sharply with the situation in general relativity, where analogous instabilities arise at macroscopic separations from the inner horizon.
In this regime, the classical description of mass inflation is no longer reliable.

The thin-shell analysis based on the DHBI junction conditions supports this conclusion. The effective amplification factor for the interior mass depends quadratically on perturbative shell parameters, and the inflationary divergence appears only in an unphysical limit in which both the shell thickness and their mutual separation fall below the fundamental length scale of the theory. This indicates that the standard classical picture of runaway mass growth is unlikely to be realized in regular QTG black holes, at least within the class of models satisfying the structural assumptions adopted here.

These results have several conceptual implications. First, the apparent suppression of mass inflation supports the viability of regular black holes with bounded-curvature cores as self-consistent solutions of modified gravity. Second, the effective stability of the inner horizon motivates a reassessment of the global causal structure of such spacetimes, including questions of determinism and the extent to which the Cauchy horizon can be regarded as a boundary of predictability. Finally, our analysis calls for a more systematic classification of regular black hole geometries based on their near-horizon response to perturbations.

There remain, however, several open questions. Our treatment relies on idealized spherical shells and does not account for generic matter fields, backreaction beyond the junction formalism, or nonspherical perturbations. Extending the analysis to fully dynamical scenarios---for example through numerical relativity in QTG or perturbative evolution on fixed backgrounds---would provide a more decisive assessment of interior stability. It would also be of interest to understand whether an analogous suppression of inflation occurs in rotating or charged regular black hole solutions, where additional structure may alter the near-horizon balance of ingoing and outgoing fluxes.

In summary, the evidence presented here indicates that mass inflation is not an unavoidable feature of regular black holes in quasitopological gravity. Instead, it appears to be strongly constrained and likely absent for physically realizable configurations. This behavior distinguishes QTG models from their general relativistic counterparts and highlights the role of a fundamental geometric cutoff in regulating black hole interior dynamics.

\acknowledgments

The authors thank the Natural Sciences and Engineering
Research Council of Canada and the Killam Trust for their
financial support.

\appendix

\section{Mass inflation in the charged black hole}

To illustrate the mass inflation effect in black holes in the general relativity we consider a Reissner-Nordstr\"{o}m solution describing a charged  black hole. Let $M$ and $Q$ be its mass and charge, respectively. In four-dimensional spacetime the metric is
\be
\begin{split}
&ds^2=-f dt^2+\dfrac{dr^2}{f}+r^2 d\omega^2\, ,\\
&f=1-\dfrac{2M}{r}+\dfrac{Q}{r^2}=\dfrac{(r-r_+)(r-r_-)}{r^2}\, ,\\
&r_{\pm}=M\pm\sqrt{M^2-Q^2}\, ,\\
&\kappa_-=-\dfrac{\sqrt{M^2-Q^2}}{M-\sqrt{M^2-Q^2}}\, .
\end{split}
\ee
Here $\kappa_-$ is (negative) surface gravity at the inner horizon $r=r_-$.

For null spheres intersecting  close to the horizon at $r=r_-(1+x)$, one finds
\be
f_B\approx 2\kappa_- x\, .
\ee
If the masses of the intersecting spheres are $\Delta M_C$ and  $\Delta M_C$ then
\be
\begin{split}
&f_C\approx f_B-\dfrac{2\Delta M_C}{r_-}\, ,\\
&f_B\approx f_B-\dfrac{2\Delta M_D}{r_-}\, .
\end{split}
\ee
Using these results, one finds
\be
f_A=\dfrac{f_C f_D}{f_A}\approx
\dfrac{2\Delta M_C\Delta M_D}{\kappa_- r_-^2}\, .
\ee
Hence the criterium $|f_A|\sim 1$ meaning that the metric in the sector $A$ at the intersection of shells significantly shifts from its near-horizon value happens  when
\be \n{RNrr}
r-r_-=r_- x\approx \dfrac{2\Delta M_C\Delta M_D}{\sqrt{M^2-Q^2}}\, .
\ee




%

\end{document}